\documentclass[aps,pre,twocolumn,preprintnumbers,superscriptaddress]{revtex4}
\usepackage{color}
\usepackage{multirow}
\usepackage[normalem]{ulem}
\usepackage{amsmath}
\usepackage{enumerate}
\usepackage{amsfonts}
\usepackage{epsfig}
\usepackage{graphicx}
\usepackage[caption=false]{subfig}
\usepackage{floatrow}
\usepackage{subfig}
\usepackage{hyperref}
\newcommand{\be}{\begin{equation}}
\newcommand{\ee}{\end{equation}}
\newcommand{\bea}{\begin{eqnarray}}
\newcommand{\eea}{\end{eqnarray}}

\newcommand{\SYdel}[1]{}

\newcommand{\es}[2] {\begin{equation} \label{#1} \begin{split} #2 \end{split} \end{equation}}

\def\le{\left}
\def\ri{\right}

\def\lssg{Wahnstr\"{o}m }

\begin{document}

\title{Linking dynamical heterogeneity to static amorphous order}

\preprint{MIT-CTP-4496}

\author{Patrick Charbonneau}
\affiliation{Department of Chemistry, Duke University,
Durham, North Carolina 27708, USA}
\affiliation{Department of Physics, Duke University,
Durham, North Carolina 27708, USA}

\author{Ethan Dyer}
\affiliation{Center for Theoretical Physics, Massachusetts Institute of Technology,
Cambridge, Massachusetts 02139, USA}
\affiliation{Stanford Institute for Theoretical Physics, Stanford University,
Stanford, California 94305, USA}

\author{Jaehoon Lee}
\affiliation{Center for Theoretical Physics, Massachusetts Institute of Technology,
Cambridge, Massachusetts 02139, USA}
\affiliation{Department of Physics and Astronomy, University of British Columbia,
Vancouver, British Columbia V6T 1W9, Canada}

\author{Sho Yaida}
\affiliation{Center for Theoretical Physics, Massachusetts Institute of Technology,
Cambridge, Massachusetts 02139, USA}
\affiliation{Department of Chemistry, Duke University,
Durham, North Carolina 27708, USA}


\begin{abstract}
Glass-forming liquids grow dramatically sluggish upon cooling. This slowdown has long been thought to be accompanied by a growing correlation length. Characteristic dynamical and static length scales, however, have been observed to grow at different rates, which perplexes the relationship between the two and with the slowdown. Here, we show the existence of a direct link between dynamical sluggishness and static point-to-set correlations, holding at the local level as we probe different environments within a liquid. This link, which is stronger and more general than that observed with locally preferred structures, suggests the existence of an intimate relationship between structure and dynamics in a broader range of glass-forming liquids than previously thought.
\end{abstract}
\maketitle

\newpage

\section{Introduction}
Despite the widespread use of glasses, a microscopic understanding of their formation remains elusive.
Any such description must incorporate the most salient characteristics of glass formation: the shear viscosity and the structural relaxation time grow by many orders of magnitude when cooling a liquid over a modest range of temperatures, without any obvious structural change.
The ubiquity of glass formation has led to a search for a universal description of the process, often invoking an underlying diverging length scale that accompanies the marked growth in relaxation time.
Finding evidence for such length scale in traditional observables has proved difficult~\cite{ENG91,SK00,KT08,BB11}, and the challenge has spurred the search for an even richer set of observables. Two main families of attempts at identifying a growing length scale have been pursued: one focusing on dynamics and another focusing on statics.

On the dynamical side, there has been significant progress identifying a growing length, culminating in the characterization of spatially correlated particle motion on a characteristic scale~\cite{AG65,Ediger00, BBBCS11}.
This dynamical heterogeneity corresponds to the system transiently fragmenting into regions with high and low mobility, before homogenizing anew over long timescales. The average size of these dynamically heterogeneous domains, $\xi_{\rm dyn}$, clearly grows as temperature is lowered~\cite{Ediger00,GSSG01,LSSNG02,LSSG03,VG04,KDS10,FS10,FZS11,BBBCS11,FS13,FSS14,ABMBBLLTWL16}.

On the static side, the situation is more muddled. Many candidate length scales have been proposed with various degrees of success~\cite{KTW89,BB04,MS06,CP07,MDIO10,TKSW10,KL11,DWR12,XSA12}.
A long-time favorite measures the growth of locally preferred structures (LPS) in certain glass formers~\cite{Frank52,SNR83}.
For example in the \lssg liquid~\cite{Wahnstrom91}, dynamical slowdown is accompanied by a marked increase of icosahedral ordering~\cite{CP07}.
A key drawback of this approach, however, is that LPS and their relation to dynamical slowdown are highly system-dependent~\cite{HCIR14}.
Point-to-set (PTS) correlations, by contrast, provide a system-independent approach~\cite{BB04,MS06}.
As long as one considers coarse-grained degrees of freedom that get frozen upon glass formation, \textit{i.e.}, particle positions or their local bond-orientations, PTS correlations universally pick up a growing amorphous order~\cite{YBCT16}.
In other words, PTS correlations detect the rarefaction of metastable states in local free-energy landscapes, the number of which is expected to sharply decrease upon approaching a putative entropy crisis, such as at the extrapolated Kauzmann temperature, $T_\mathrm{K}$~\cite{Kauzmann48}.
Encouragingly, the PTS correlation length has been found to grow more than simpler static lengths in various glass-forming liquids~\cite{CGV07,BBCGV08,SL11,HMR12,CCT12,CB12,JB12,KVB12,BICtest12,BK12,CCT13,BKP13,CT13,KB13,GTCGV13,HBKR14,OKIM15,BCY16}.

While both the dynamic, $\xi_{\rm dyn}$, and the static PTS, $\xi_{\rm PTS}$, lengths seem to universally increase upon cooling, the relation between the two remains unclear.
In the numerically accessible regime over which they have been determined they grow at different rates, the former more rapidly than the latter~\cite{CT13,BCY16}.
One might thus be tempted to dispose of PTS correlations as irrelevant to glassy dynamics -- as is sometimes invoked to motivate certain dynamical models of the glass transition~\cite{GC03}.
The lack of a linear relation between the two types of length scales, however, does not necessarily imply the absence of any relation.
There is indeed a recurring observation that regions with a higher concentration of LPS tend to relax more slowly than the rest of the system~\cite{DSZ02,ST06,PSDH10,TKSW10,LT12,MERWT13,METR13,RMDP14,RW15}.
This link seems to persist even in glass formers that display a clear divergence between the dynamical and static LPS lengths, notably for the \lssg liquid~\cite{HCIR14}.
The problem, of course, is that this relationship is highly system-dependent, like the LPS themselves.
Consequently, it does not obviously hold in the absence of aggregating LPS, as is for instance the case for the Kob-Andersen liquid~\cite{HCIR14}.
Such an approach, however elaborate, thus offers little hope of attaining a universal theory of the glassy slowdown.

Here through a different route, we provide evidence for a system-independent, universal link between statics and dynamics in glass-forming liquids.
By focusing on the general amorphous order rather than on the system-specific LPS order, we go beyond prior efforts, finding a strong link in the Kob-Andersen liquid, for which a LPS-based approach yields only a weak link~\cite{HCIR14}. 
More precisely, we find that the link, quantified in terms of the Spearman coefficient, is most manifest when PTS correlations are measured with cavities of size $R\sim\xi_{\rm PTS}$ at low temperatures.

Before going any further, however, let us say a few words about amorphous ordering. The notion was introduced by Biroli and Bouchaud~\cite{BB04}--and formalized by Montanari and Semerjian~\cite{MS06}--so as to describe the physical necessity that a liquid develops some sort of structural order as the number of possible equilibrium configurations decreases, \textit{i.e.}, as temperature decreases or density increases. This order, however, has generically nothing to do with structures that give rise to crystalline ordering, hence its amorphous nature. One might expect amorphous order to be controlled by disordered constraints, such as the randomness of the couplings in spin glasses, that have no equivalent in liquids. 
The ingenuity of point-to-set correlations as applied to systems without such quenched disorder lies in introducing a self-induced disorder through pinning a subset of particles, and in then measuring its influence on the rest of the system. The existence of nontrivial, amorphous order can thus be assessed even in simple glass-forming liquids.

The rest of this article is organized as follows.
After detailing our simulation protocols in Sec.~\ref{simulations}, we define the overlap field in Sec.~\ref{overlap}, which measures the local similarity between two configurations.
We use this observable to characterize local sluggishness, make precise the notion of the local amorphous order, and relate the two in Sec.~\ref{results}.
We conclude and discuss possible future directions in Sec.~\ref{conclusion}.

\section{Model and simulation details}
\label{simulations}
We simulate the canonical glass former originally proposed by Kob and Andersen in $d=3$ spatial dimensions~\cite{KA94, KA95}.
The liquid contains two particle species, denoted by $A$ and $B$, with equal mass, $m$, interacting via a Lennard-Jones potential,
\es{LJ}{
V_{ab} ( r) = 4 \varepsilon_{ab} \le[ \le(\frac{\sigma_{ab}}{r}\ri)^{12} - \le(\frac{\sigma_{ab}}{r}\ri)^6 \ri]\ ,}
where $a,b \in \{A, B\}$, and the parameters satisfy $\epsilon_{AB} / \varepsilon_{AA}= 1.5$, $\varepsilon_{BB} / \varepsilon_{AA} = 0.5$, $\sigma_{AB} / \sigma_{AA} = 0.8$, and $\sigma_{BB}/\sigma_{AA} =0.88$.
The interaction potential is truncated and shifted at $r^{\rm cut}_{ab}=2.5 \sigma_{ab}$.
The relative number of particles is $N_A:N_B = 4:1$ and the overall number density is $\rho=1.2\sigma_{AA}^{-3}$.
Length, temperature $T$, and time $t$ are reported in standard Lennard-Jones dimensionless units set by $\sigma_{AA}$, $\varepsilon_{AA}/k_{\rm B}$, and $(m \sigma_{AA}^{2} /  \varepsilon_{AA} )^{1/2}$, respectively.

We study samples with $N=135,000$ particles in a periodic cubic box using molecular dynamics simulations carried out with LAMMPS~\cite{Lammps},  accelerated by the GPU package~\cite{LammpsGPU1,LammpsGPU2} with mixed precision~\cite{GPUmixed}.
In order to preserve the numerical stability of the algorithm, we use the velocity-Verlet time integration scheme with an integration timestep $dt=0.005$, and a neighbor list with skin depth $0.3$, updated every $20$ molecular dynamics steps.
Simulations are performed in the canonical, constant $NVT$, ensemble, using the Nos{\'e}-Hoover thermostat \cite{Nose,Hoover} with damp time $1.0$.
Drift in the center of mass is prevented by keeping its position constant at every integration step.

We prepare samples by starting from an equilibrium configuration at $T=2.000$ and sequentially equilibrating at $T=1.000,0.800,0.600,0.550,0.510,0.485, 0.465$ and finally $0.450$, which is close to the mode-coupling temperature, $T_{\rm MCT}=0.435$~\cite{KA94, KA95}.
Specifically, we cool a sample from one temperature to the next with a slow cooling rate, $ \frac{dT}{dt} \lesssim 10^{-4}\frac{1}{ \tau_{\alpha} (T)}$, followed by an equilibration run for time   $t_{\rm equi}$ with $120\tau_{\alpha}<t_{\rm equi}<160\tau_{\alpha}$ (except that $t_{\rm equi}\approx280\tau_{\alpha}$ at $T=1.000$), where the structural relaxation time $\tau_{\alpha}$ is defined below.

Throughout the simulation, the total energy and pressure are monitored in order to detect any hint of numerical instability or crystallization.
As a further check, a second, statistically independent sample with different initial velocities is used to validate the bulk measurements, \textit{i.e.}, energy, pressure, and dynamical correlation functions.
Note that for $T<0.450$, crystallization is found to interfere with dynamical measurements~\cite{TPSD09} for our choice of dynamical protocol, which sets the lower $T$ for our study.

In order to measure bulk liquid properties, equilibrated samples are evolved at each temperature for time $t_{\rm prod}=t_{\rm equi}$, while keeping the system coupled to the thermostat, and $n=1000$ equally-spaced configurations are recorded, \textit{i.e.}, they are separated by $t_{\rm rec}=t_{\rm prod}/n$.
We denote each configuration ${\bf X}_{(s)}=\le\{({\bf x})^a_i(t=s t_{\rm rec})\ri\}$, where $s=0,...,n-1$ is the snapshot number, $a\in\{A,B\}$, and $i=1,...,N_a$ the particle number.

\section{Overlap field and bulk dynamics}\label{overlap}
The dynamics and growing amorphous order of glass-forming liquids can be characterized in various ways.
A common dynamical measure is the decay of the intermediate scattering function, which has the advantage of being observable through light scattering experiments~\cite{HM86}.
Here, we use an overlap field, $q({\bf r})$, because it allows us to naturally compare the liquid dynamics against PTS correlations, as defined in Sect.~\ref{results}.
The usage of the continuous overlap field also smoothens out the short-range oscillation of the dynamical correlation functions due to cage sampling, making it possible to extract the dynamical length through a relatively simple fitting.

\subsection{Overlap field}
We denote a pair of configurations by ${\bf X}=\le\{{\bf x}^a_i\ri\}$ and ${\bf Y}=\le\{{\bf y}^a_i\ri\}$.
For each particle ${\bf x}_i^a$, we find the nearest particle ${\bf y}_{i_{\rm nn}}^a$ of the same species, and assign an overlap value $q_{{\bf X; Y}} \le({\bf x}_i^a\ri)\equiv w\le(\big|{\bf x}_i^a-{\bf y}_{i_{\rm nn}}^a\big|\ri)$, where
\be
w(z)\equiv \exp\le[-\le(\frac{z}{b}\ri)^2\ri]\, 
\ee
with $b=0.2$.
This procedure defines overlap values $q_{\bf X; Y} \le({\bf x}_i^a\ri)$ at scattered points $\le\{{\bf x}_i^a\ri\}$.
We define $q_{\bf X; Y} \le({\bf r}\ri)$ to be a continuous function, linearly interpolating these overlap values within Delaunay simplices spanned by the scattered points.
Similarly, we define $q_{\bf Y; X} \le({\bf r}\ri)$ and set
\be
q_{\bf X,Y}\le({\bf r}\ri)\equiv \frac{1}{2}\le\{q_{\bf X; Y}\le({\bf r}\ri)+q_{\bf Y; X}\le({\bf r}\ri)\ri\}\, .
\ee
This overlap field quantifies the local similarity around a point ${\bf r}$ between two configurations, ${\bf X}$ and ${\bf Y}$.

\subsection{Bulk relaxation}
\begin{figure*}[t]
\centering{
\sidesubfloat[]{\includegraphics[width=0.27\textwidth]{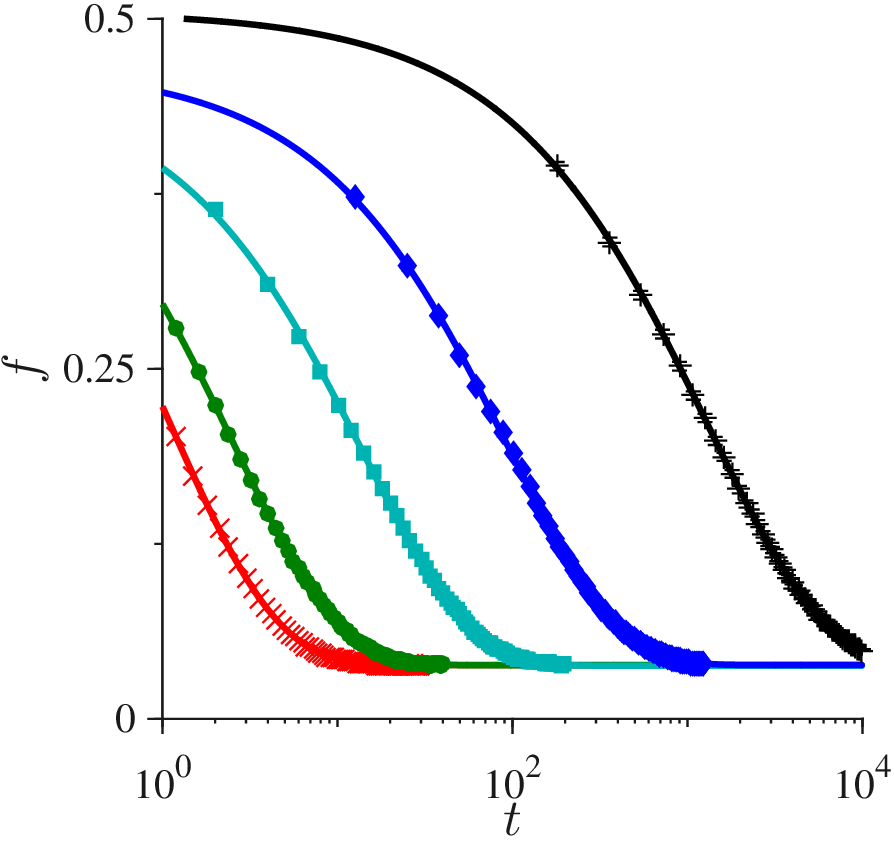}}\quad%
\sidesubfloat[]{\includegraphics[width=0.27\textwidth]{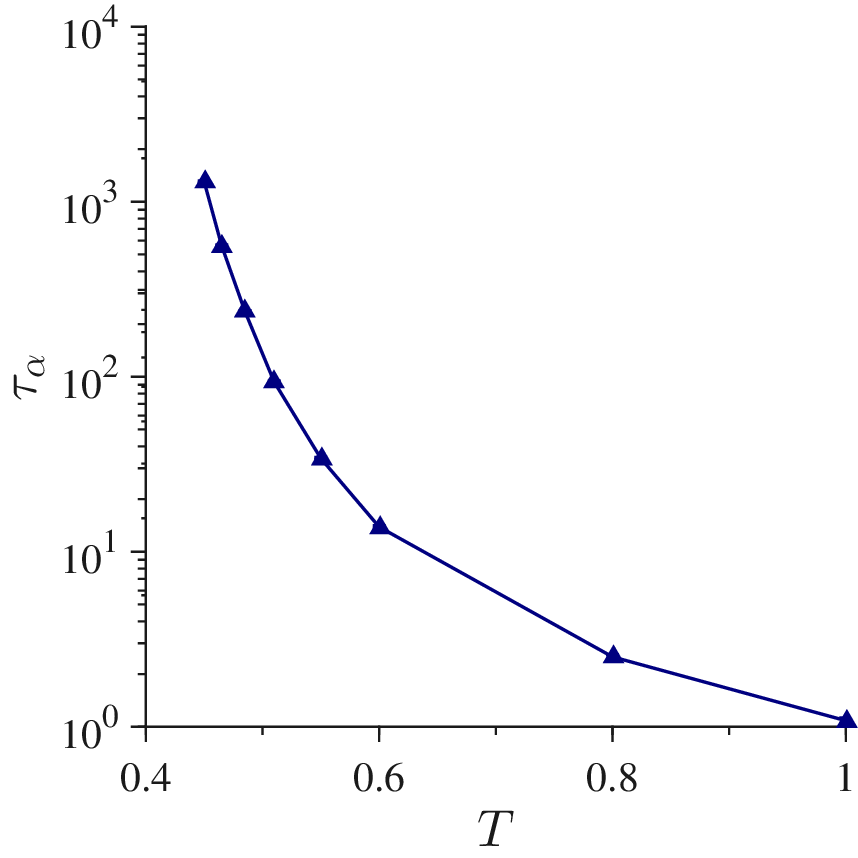}}
}
\caption{Structural relaxation of the Kob-Andersen liquid seen through the dynamical overlap field. (a) Autocorrelation function $f(t)$ at temperatures $T = 1.000$ (red-cross), $0.800$ (green-circle), $0.600$ (cyan-square), $0.510$ (blue-diamond), and $0.450$ (black-plus). Solid lines are stretched exponential fits to Eq.~\eqref{eq:fit}, where $f(t)$ asymptotes to $f_{\infty}\approx0.04$ at long times. (b) The structural relaxation time, $\tau_{\alpha}$, rapidly grows as temperature decreases.}
\label{Auto}
\end{figure*}

Using this definition, we can cast the dynamical slowdown of glass-forming liquids in the language of the overlap field.
When a liquid is sufficiently cooled, each constituent particle is caged by its neighbors and vibrates around its itinerant center.
At short times, $t\ll\tau_{\alpha}$, an initial configuration, ${\bf X}_0$, and its time-evolved configuration, ${\bf X}_t$, look quite similar.
In other words, the dynamical overlap field
\be
q_{{\bf X}_0}\le({\bf r};t\ri)\equiv q_{{\bf X}_0,{\bf X}_t}\le({\bf r}\ri)
\ee
takes high values nearly everywhere.
At long times, $t\gg\tau_{\alpha}$, the liquid explores phase space and thus loses its similarity to ${\bf X}_0$, resulting in low overlap values.
The crossover from similar to dissimilar structures defines $\tau_\alpha$.
More precisely, we first measure the autocorrelation function
\be\label{auto}
f(t)\equiv \langle q_{{\bf X}_0}\le({\bf r}_0;t\ri) \rangle \, ,
\ee
where $\langle\ldots\rangle$ indicates the thermal average over equilibrium configurations ${\bf X}_0$ and positions ${\bf r}_0$ (Fig.~\ref{Auto}a). We then perform a least-square fit to the stretched exponential function 
\be\label{eq:fit}
f_{\mathrm{fit}}(t)=A \exp\le\{-\le(\frac{t}{\tau_{\alpha}}\ri)^{\gamma}\ri\}+f_{\infty}\, 
\ee
for $t\geq1$, where the asymptotic value, $f_{\infty}$, is a static quantity that depends on the choice of the overlap function $w(r)$. The thermal average is evaluated by taking $10$ initial equilibrium configurations ${\bf X}_0=\mathbf{X}_{(s)}$ at $s=0,100,\ldots,900$ and by comparing them against the ones at time $t=k t_{\rm rec}$ for $k=1,2,\ldots,99$. The subsequent averaging over positions is implemented by randomly selecting $1000$ points in each initial configuration, associating to each point a mean value over a ball of radius $1.0$ around it through Monte Carlo integration with $10^5$ points, and then averaging over these means.  The value of $f_{\infty}$ is evaluated similarly by repeatedly taking $20000$ pairs of well-separated points in the bulk and again further averaging over mean overlap values within a ball of radius $1.0$ centered around these points. As expected, the resulting structural relaxation time quickly grows as temperature decreases (Fig.~\ref{Auto}b).

\subsection{Dynamical heterogeneity}

\begin{figure*}[t]
\centering{
\sidesubfloat[]{\includegraphics[width=0.27\textwidth]{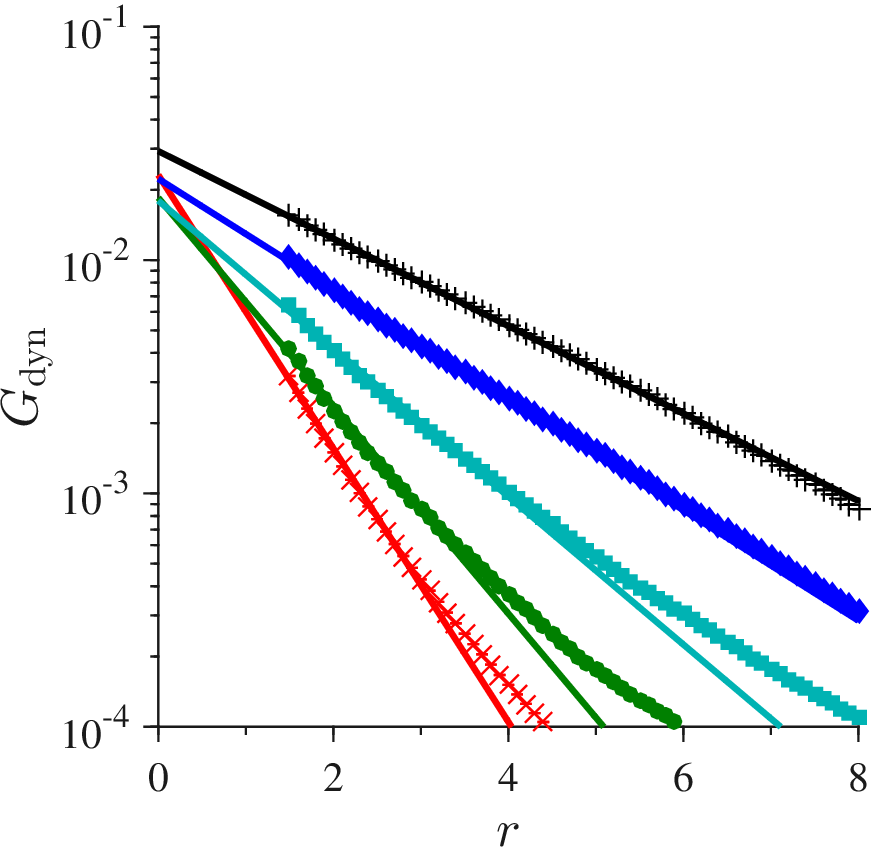}}\quad%
\sidesubfloat[]{\includegraphics[width=0.27\textwidth]{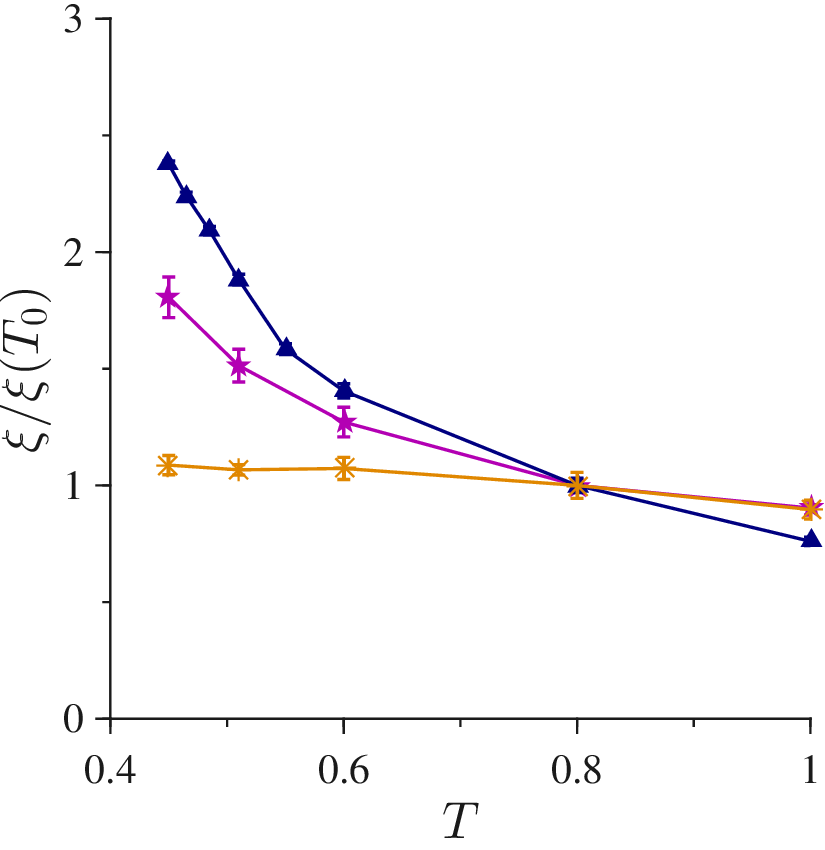}}
}
\caption{Dynamical heterogeneity in the Kob-Andersen liquid seen through the dynamical overlap field. (a) Dynamical correlation function at $t\approx\tau_{\alpha}$. Color codes are the same as in Fig.~\ref{Auto}. (b) The growing dynamical length, $\xi_{\rm dyn}$ (navy blue-triangle), is compared to the PTS length, $\xi_{\rm PTS}$ (purple-pentagram), extracted as in~Ref.~\cite{BCY16}. The length extracted from the two-point density-density static correlation function, $\xi_{2}$ (orange-asterisk), is also included for reference from Ref.~\cite{YBCT16}. Note that all are rescaled to unity at $T_0=0.8$, where the onset of glassy physics gives rise to a clear separation between PTS and two-point lengths~\cite{YBCT16}.}
\label{DynamicGrowth}
\end{figure*}

The spatial distribution of the decorrelating overlap field reveals the dynamical heterogeneity of the relaxation process, and this heterogeneity is most pronounced at $t\sim\tau_{\alpha}$, as has been broadly reported~\cite{BBBCS11}. On the $\tau_\alpha$ timescale, such a system is commonly subdivided into (a minimum of) two types of regions, dubbed fast and slow.
In fast regions, cooperative rearrangements quickly erase the similarity to the initial configurations, resulting in relatively low overlaps; in slow regions, particles have barely moved, resulting in relatively high overlaps.
The typical size of these regions, $\xi_{\rm dyn}$, is encoded in the connected dynamical correlation function
\be\label{DC}
G_{\rm dyn}\le({\bf r}; t\ri)\equiv \langle q_{{\bf X}_0}\le({\bf r}+{\bf r}_0;t\ri)q_{{\bf X}_0}\le({\bf r}_0;t\ri)\rangle - f(t)^2\, ,
\ee
where we take $t=\tau_{\alpha}$ (Fig.~\ref{DynamicGrowth}a). 
Note that this function involves two ${\bf X}_0$'s and two ${\bf X}_t$'s, and is thus closely related to standard four-point correlation functions used to characterize the size of heterogeneous domains~\cite{BBBCS11}.

In order to average the two-point components, we take $100$ initial configurations ${\bf X}_0$ at $s=0,10,\ldots,990$ and randomly choose $5000$ points ${\bf r}_0$ in each; to evaluate the product $q_{{\bf X}_0}\le({\bf r}+{\bf r}_0;t\ri)q_{{\bf X}_0}\le({\bf r}_0;t\ri)$ at radius $r$, we further average over orientation by taking $10^3$ points $\mathbf{r}$ uniformly distributed over the sphere of radius $r$. To estimate the dynamical correlation function at $t=\tau_{\alpha}$, we choose $k_{\alpha}$ such that $k_{\alpha}t_{\rm rec}$ is closest to $\tau_{\alpha}$ among all integers. In order to extract a dynamical length, we finally fit the radial decay of this function to an exponential form
\begin{equation}
G_\mathrm{fit}\le(r\ri)=B \exp\le(-\frac{r}{\xi_{\rm dyn}}\ri)\, ,
\end{equation}
where $B$ and $\xi_{\rm dyn}$ are fitting constants.
Note that in order to reduce statistical noise in the asymptotic tail of connected dynamical correlation function, $f(k_{\alpha}t_{\rm rec})$ is evaluated more accurately than before, averaging over $100$ initial configurations and $5000$ randomly chosen balls of radius $1.0$ within each of those.

As expected from earlier studies~\cite{Ediger00,GSSG01,LSSNG02,LSSG03,VG04,KDS10,FS10,FZS11,BBBCS11,FS13,FSS14}, the range of dynamical correlations grows as temperature decreases, and does so more rapidly than the PTS correlation length, $\xi_{\rm PTS}$ (from Ref.~\cite{BCY16}), all the while the two-point density-density correlation length, $\xi_2$ (from Ref.~\cite{YBCT16}), remains roughly constant (Fig.~\ref{DynamicGrowth}b).
The divergence between $\xi_{\rm dyn}$ and $\xi_{\rm PTS}$ was first noted in hard-sphere glass formers within the random pinning protocol~\cite{CT13}, and our results confirm that the phenomenon is fairly generic.

\section{Local observables}\label{results}
Given a point $\mathbf{r}_0$ within a configuration $\mathbf{X}_0$, we qualitatively expect that the more extended the amorphous order, the more cooperative the rearrangements, and the more sluggish the particle evolution.
In this section, we demonstrate this expectation quantitatively.
We first explain how to characterize local dynamics and the extent of amorphous order using the overlap field, and then observe a direct link between the two quantities.

\subsection{Local dynamics}
\label{LD}
In order to quantify the local mobility at point $\mathbf{r}_0$, one can probe the dynamical overlap field around it. To reduce irrelevant vibrational noise, however, one must first average locally,
\be
{q}_{\rm dyn}(\mathbf{r}_0; t)\equiv \frac{3}{4\pi }\int_{|{\bf r}|<1}\mathrm{d}{\bf r}\ q_{\mathbf{X}_0,\mathbf{X}_t}\le({\bf r}+\mathbf{r}_0\ri)\, ,
\ee
where the integral is evaluated by Monte Carlo integration with $10^5$ points, and then further average over time
\be\label{TA}
\bar{q}_{\rm dyn}(\mathbf{r}_0)\equiv\frac{1}{\tau_{\alpha}/2}\int_0^{\tau_{\alpha}/2} \mathrm{d}t {q}_{\rm dyn}(\mathbf{r}_0; t)\, ,
\ee
which we here implement by averaging over configurations between times $0$ and $\tau_{\alpha}/2$.
Note that our sampling time, $t_{\rm rec}\sim0.1\tau_{\alpha}$, is rather coarse and thus a large residual thermal noise remains in our evaluation of time averages, especially at high temperatures.
As temperature decreases, however, the signal to noise ratio increases and we are then able to pick up the correlation between statics and dynamics.

In a region with high mobility the dynamical overlap decays relatively quickly with time, resulting in a low $\bar{q}_{\rm dyn}$, while in a region of low mobility it retains a large value for a relatively long time, resulting in a high $\bar{q}_{\rm dyn}$.
Local sluggishness can thus be quantified by considering $\bar{q}_{\rm dyn}(\mathbf{r}_0)$.

\begin{figure*}[t]
\centering{
\sidesubfloat[]{\includegraphics[width=0.3\textwidth]{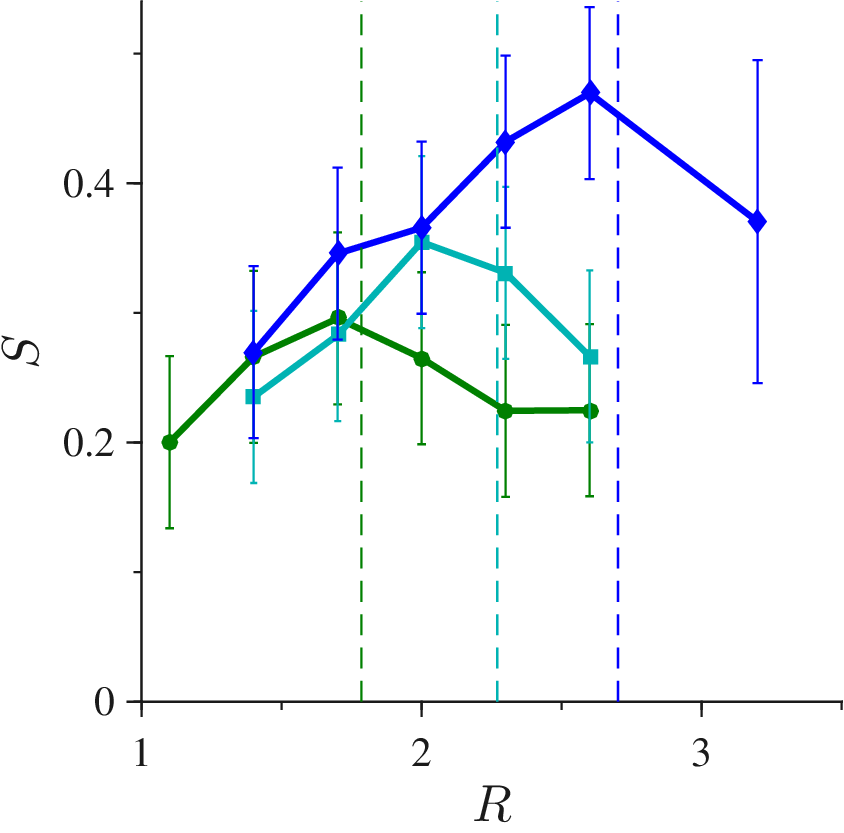}}\quad%
\hspace{+0.3in}\sidesubfloat[]{\includegraphics[width=0.4\textwidth]{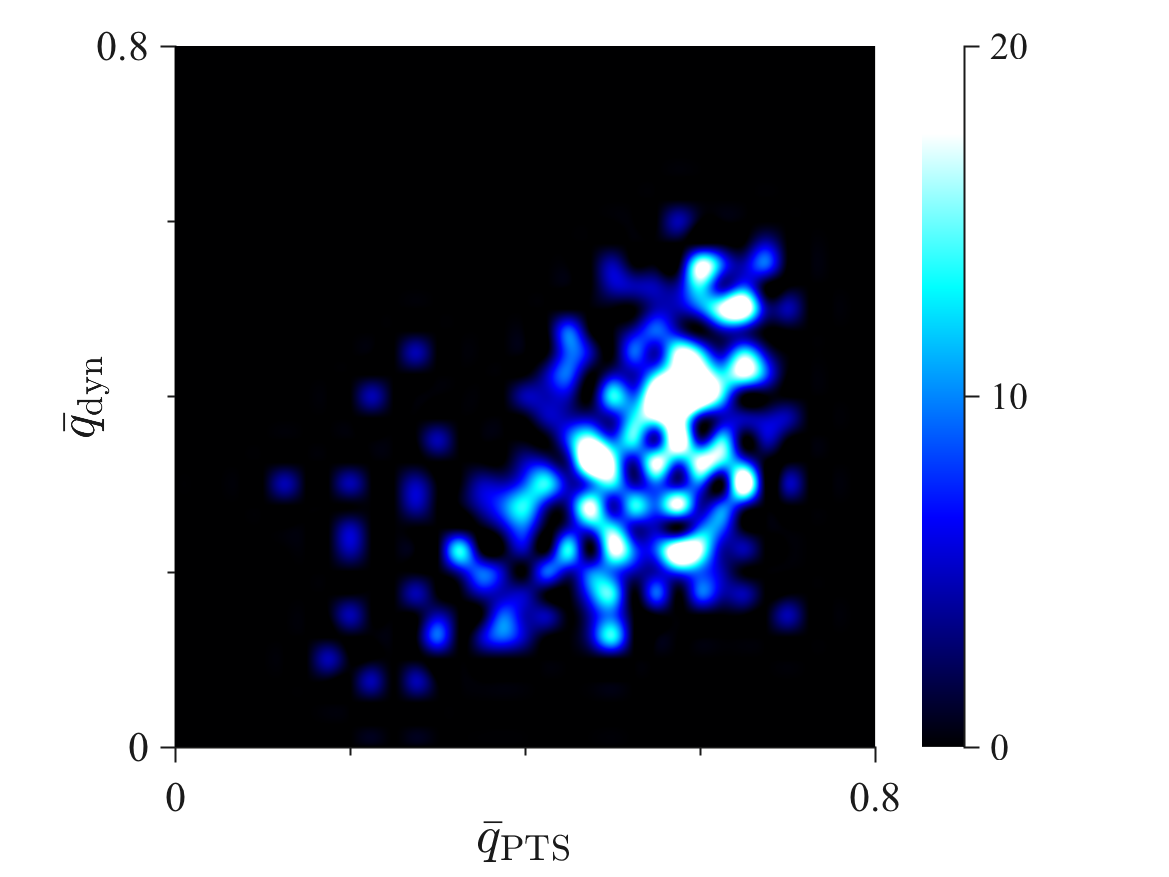}}}
\caption{ (a) Spearman coefficient $S$ as a function of cavity radius $R$ for $T=0.8,0.6,0.51$. Color code is the same as in Fig.~\ref{Auto} and error bars are $95\%$ confidence intervals, \textit{i.e.}, $\pm2\tilde{\sigma}$. The results are consistent with the Spearman coefficient taking its highest value around $0.8 \xi_{\rm PTS}$, marked by dotted lines.
(b) Probability distribution of $(\bar{q}_{\rm PTS}, \bar{q}_{\rm dyn})$ for the Kob-Andersen liquid at $T=0.51$ and $R=2.6$.
Here, $\bar{q}_{\rm PTS}$ characterizes the spatial extent of local amorphous order while $\bar{q}_{\rm dyn}$ characterizes the local sluggishness. The high value of the associated Spearman coefficient, $S=0.47$, indicates a roughly monotonic relation between these two quantities.}
\label{Peak}
\end{figure*}

\subsection{Local PTS correlations}\label{PTS}
PTS correlations characterize the local free-energy landscape around a point $\mathbf{r}_0$, within a configuration $\mathbf{X}_0$.  In order to measure this purely static correlation, we pin particles outside a cavity of radius $R$ centered at $\mathbf{r}_0$, sample new configurations $\widetilde{\mathbf{X}}$ within that cavity, and measure the statistics of the overlap field $q_{\mathbf{X}_0,\widetilde{\mathbf{X}}}\le({\bf r}\ri)$.
Equilibration inside the cavity is carried out following the parallel-tempering scheme presented in Ref.~\cite{BCY16} and using the same parameters, except for $T=0.8$ and $R=1.1$, which is new here. In this case, temperatures and shrinking parameters for replicas, $\le\{(T_a,\lambda_a)\ri\}_{a=1,...,n}$, satisfy the linear relation
$\frac{T_a-T_1}{T_{\rm dec}-T_1}=\frac{\lambda_a-\lambda_1}{\lambda_{\rm dec}-\lambda_1}$,
with $T_{\rm dec}=1.00$, $\lambda_{\rm dec}=0.80$, and  $\le(\lambda_1,\lambda_2,\lambda_3,\lambda_4,\lambda_5,\lambda_6,\lambda_7,\lambda_8\ri)=\le(1.000,0.955,0.910,0.865,0.820,0.775,0.730,0.685\ri)$; see Ref.~\cite{BCY16} for methodological details.
For simplicity, we here focus on average values near the cavity core,
\be
\bar{q}_{\rm PTS}(\mathbf{r}_0; R)\equiv \frac{3}{4\pi}\int_{|{\bf r}|<1}\mathrm{d}{\bf r}\ \langle q_{\mathbf{X}_0,\widetilde{\mathbf{X}}}\le({\bf r}+\mathbf{r}_0\ri)\rangle_{R}\, ,
\ee
where the conditional thermal average $\langle\ldots\rangle_{R}$ denotes averaging over constrained equilibrium configurations $\widetilde{\mathbf{X}}$, under the influence of the pinned particles outside a cavity of size $R$, which acts as an effective quenched disorder. 
The integral is evaluated by Monte Carlo integration with $10^5$ points. Note that although we sample configurations $10$ times less frequently than in Ref.~\cite{BCY16}, the averaging is more than sufficient to properly evaluate core overlap values.
In order to obtain good statistics for later analysis, we also sample $350$ distinct $\mathbf{r}_0$ (100 for $T=0.51$ and $R=3.2$), which is markedly more than Ref.~\cite{BCY16}.
Results are obtained for $T=0.80$, $0.60$, and $0.51$, but in order to keep the computational cost within reason the computations are carried out only with cavity sizes $R\leq2.6$ (plus $R=3.2$ for $T=0.51$).
Even within this reduced $R$ regime, we detect a significant correlation between the static amorphous order and dynamical sluggishness.

When simulations are carried out for small $R\ll\xi_{\rm PTS}$, pinning is so constraining that $\mathbf{X}_0$ and $\widetilde{\mathbf{X}}$ are very similar, resulting in high overlap with high probability.
By contrast, for large $R\gg\xi_{\rm PTS}$, the impact of the boundary is negligible and thus two configurations are essentially statistically independent near the core, resulting in low overlap with high probability.
In the intermediate regime $R\sim\xi_{\rm PTS}$, disorder fluctuations in the local free-energy landscape emerge.
Specifically, when the probing cavity size is near $R\sim\xi_{\rm PTS}$, some region contains a single minimum, while some other region holds more than a few.
In general we expect that the fewer metastable states the local landscape holds, the broader the amorphous order extends around $\mathbf{r}_0$, the farther inside the cavity nonperturbative boundary effects propagate. As a result, $\bar{q}_{\rm PTS}(\mathbf{r}_0; R)$ then decays less quickly.
Examining $\bar{q}_{\rm PTS}(\mathbf{r}_0; R)$ with $R\sim\xi_{\rm PTS}$, we can thus quantitatively diagnose the relative spatial extent of amorphous order in a given local environment.
In this language, the PTS length scale corresponds to the average spatial extent of the fluctuating amorphous order.

\subsection{Local link between dynamics and static PTS}
We now have the necessary tools to analyze the relationship between the dynamical observable, $\bar{q}_{\rm dyn}(\mathbf{r}_0)$, characterizing local sluggishness, and the static observable, $\bar{q}_{\rm PTS}(\mathbf{r}_0; R\sim\xi_{\rm PTS})$, characterizing local amorphous order.
Based on the results of earlier studies, we do not expect a perfect one-to-one linear mapping between the two, but rather a roughly monotonic relation.
A standard measure for quantifying such a relationship is the Spearman's rank correlation coefficient, $S$, which assesses correlation between two quantities in the ranking space~\cite{Spearman1904}.
Specifically, for a set of quantities $\mathcal{W}=\le\{W_m\ri\}_{m=1,\ldots,M}$, we define the rank $w_m={\rm Rank}\le(\mathcal{W}; m\ri)$ such that $W_m$ is the $w_m$-th highest value in the set.
Then, given $M$ observations of two quantities, $\mathcal{W}$ and $\mathcal{Z}$, we define
\be
S=1-\frac{\sum_{m=1}^M6\le\{{\rm Rank}\le(\mathcal{W}; m\ri)-{\rm Rank}\le(\mathcal{Z}; m\ri)\ri\}^2}{M(M^2-1)}\, .
\ee
This coefficient attains $|S|=1$ for a perfect monotonic relation, and asymptotes to $S=0$ for large uncorrelated dataset. Its standard error is given by $\tilde{\sigma}=0.6325/\sqrt{M-1}$.

To get an idea of the typical scale of this quantity, consider the results for the \lssg liquid~\cite{HCIR14}. Hocky \emph{et al.} found $|S|=0.62$ between local mobility and LPS, and $S=0.50$ between LPS and PTS.
Combined, these results hint at a correlation between local mobility and PTS correlation, although the corresponding $S$ was not reported.
For the Kob-Andersen liquid, the same authors found instead $|S|=0.22$ between local mobility and LPS at $T=0.45$, thus concluding that LPS is not as good a dynamical predictor for this system as it is for the \lssg liquid. Again, this distinction likely reflects the system-dependent nature of the predicting power of LPS. 

Consider now how the general amorphous order fares as a dynamical predictor by evaluating Spearman coefficients between $\bar{q}_{\rm PTS}(\mathbf{r}_0; R)$ and $\bar{q}_{\rm dyn}(\mathbf{r}_0)$ (Fig.~\ref{Peak}).
The relatively high $S$ at $T=0.51$ suggests a correlation between the static amorphous order and dynamical sluggishness that is stronger than that provided by LPS order at an even lower temperature $T=0.45$. Interestingly, the coefficients seem to be highest when $R\sim \xi_{\rm PTS}$, reaching $S=0.47(8)$ at $T=0.51$, which is consistent with the expectation that local fluctuations in the extent of the amorphous order is most saliently manifested in the PTS correlations at this scale. We thus expect that probing PTS overlap closer to $R\sim\xi_{\rm PTS}$ at other $T$ to also maximize correlation, but further study is needed to confirm the effect.
Note that our rather coarse bulk-sampling time (mentioned in Sec.~\ref{LD}) results in a large residual thermal noise in our evaluation of the dynamical overlap, especially at high temperatures (not shown).
We thus also expect the link we detect here to become even stronger for improved time-averaging of the dynamical overlap.

\section{Conclusion}\label{conclusion}
We have shown the existence of a direct positive correlation between local dynamical heterogeneity and local static PTS correlations for the Kob-Andersen liquid, despite the divergence between the two associated length scales.
This result is particularly compelling because PTS observables, unlike LPS, provide a system-independent way of linking dynamics to statics, working even in the case the latter measure fails to detect any growing order.
This robustness is accomplished by appealing to the following observation holding at the global level as glass formers probe different dynamical regimes:
the fewer metastable states the free-energy landscape holds, the farther the amorphous order extends, the more sluggishly particles move.
Our work indicates that this glassy signature holds also at the local level, within a liquid at fixed temperature as we probe different local environments.
Of course, whether of not this hypothesis \textit{universally} holds for all glass-forming liquids should be extensively tested beyond Lennard-Jones classes, notably for harmonic particles (for which the link between dynamics and LPS is virtually nonexistent~\cite{HCIR14}) and for hard spheres.
Given the unfailing observations of the glassy signature of these systems at the global level, there is a reasonable likelihood for this observation to be robust at the local level as well.

After disclosing the local link between statics and dynamics, one may still wonder why their associated lengths grow at different rates as temperature decreases.
One insight from PTS measurements is that the free-energy barriers between states are prohibitively high at the PTS scale, whereat a few distinct states become accessible. These barriers are so pronounced that extensive use of parallel-tempering is needed to overcome them within a reasonable simulation time~\cite{BCY16}.
The corresponding relaxation paths are therefore most likely not sampled in the standard bulk dynamics.
Rather, the deviation of two length scales suggests that,  as temperature decreases, one needs to go increasingly longer distances in order to access paths that enable collective motion.
The observed local link indicates that local fluctuations in the amorphous order still govern the distance at which this dynamically viable regime is reached in each region.
In order to delineate the basic physics, it might be more productive to first consider this behavior in a simple schematic model, like the one proposed in Ref.~\cite{TK95} to explain the breakdown of the Stokes-Einstein relation.
Another route might be to systematically consider the microscopic origin of the Stokes-Einstein breakdown as one approaches the mean-field regime~\cite{CCJPZ13}, as proposed in Ref.~\cite{CJPZ14}.
\\{}\\
{\bf Acknowledgments}
We thank Allan~W.~Adams for discussions and use of computational resources, William~Detmold for introducing us to the XSEDE program, and Ludovic~Berthier, Daniel~S.~Fisher, Alan~H.~Guth, Shamit~Kachru, Steven~A.~Kivelson, John~A.~McGreevy, Maksym~N.~Serbyn, and Stephen~H.~Shenker for discussions.
We acknowledge the Texas Advanced Computing Center (TACC) at the University of Texas at Austin for providing HPC resources that have contributed to the research results reported within this paper.
We also acknowledge the Duke Compute Cluster and the Open Science Grid for computational resources.
P.C.~and S.Y.~acknowledge support from the National Science Foundation Grant No.~NSF DMR-1055586, E.D.~acknowledges support from the National Science Foundation Grant No.~NSF PHY-1316699, J.L.~acknowledges support from Samsung Scholarship, and S.Y.~acknowledges support from a JSPS Postdoctoral Fellowship for Research Abroad.
Data relevant to this work have been archived and can be accessed at http://dx.doi.org/10.7924/G8VD6WC5.

\bibliography{glassD}

\end{document}